\documentclass[twocolumn,prl,superscriptaddress]{revtex4}
\usepackage[latin9]{inputenc}
\usepackage{color}
\usepackage{graphicx}
\usepackage{tikz}
\usepackage{verbatim}
\usepackage{graphicx}
\usepackage{tikz}
\usepackage{amsmath,bm,amssymb,latexsym,galois,amsthm,euscript,dsfont}
\usepackage[all,cmtip,knot]{xy}
\usepackage{dsfont} 
\usepackage{mathrsfs}
\usepackage[unicode=true,bookmarks=true,bookmarksnumbered=false,bookmarksopen=false,breaklinks=false,pdfborder={0 0 1},backref=false,colorlinks=true]{hyperref}

\hypersetup{
    colorlinks,
    linkcolor={red!50!black},
    citecolor={blue!90!black},
    urlcolor={blue!90!black}
}

\makeatletter
\@ifundefined{textcolor}{}
{%
 \definecolor{BLACK}{gray}{0}
 \definecolor{WHITE}{gray}{1}
 \definecolor{RED}{rgb}{1,0,0}
 \definecolor{GREEN}{rgb}{0,1,0}
 \definecolor{BLUE}{rgb}{0,0,1}
 \definecolor{CYAN}{cmyk}{1,0,0,0}
 \definecolor{MAGENTA}{cmyk}{0,1,0,0}
 \definecolor{YELLOW}{cmyk}{0,0,1,0}
}

\usepackage{amsfonts}\usepackage{tabularx}\usepackage{dcolumn}\usepackage{bm}\usepackage{graphicx}\usepackage{epstopdf}

\hypersetup{urlcolor=blue}

\newcommand{\Nb}{\mathbb{N}}

\newcommand\pobs{\mathbf{p}_{\mathrm{obs}}}
\newcommand{\hobs}{\mathbf{h}_{\mathrm{obs}}}

\theoremstyle{definition}
\newtheorem{defn}{Definition}

\newtheorem{lemma}[defn]{Lemma}
\newtheorem{corollary}[defn]{Corollary}
\newtheorem*{defn*}{Definition}
\newtheorem{thm}[defn]{Theorem}
\theoremstyle{remark}
\newtheorem*{pf}{PROOF}
\newtheorem*{lemma*}{Lemma\, 2}
\newtheorem*{theorem*}{Theorem\, 3}
\newtheorem*{corollary*}{Corollary\, 4}

\makeatother

\begin{document}

\title{Entropic No-Disturbance as a Physical Principle}
\author{Zhih-Ahn Jia\footnote{Two authors are of equal contributions.}}
\email{\tt giannjia@foxmail.com}
\affiliation{Key Laboratory of Quantum Information, Chinese Academy of Sciences, School of Physics, University of Science and Technology of China, Hefei, Anhui, 230026, P.R. China}
\affiliation{Synergetic Innovation Center of Quantum Information and Quantum Physics, University of Science and Technology of China, Hefei, Anhui, 230026, P.R. China}
\author{Rui Zhai\footnotesize{*}}
\email{\tt Zhairuii@foxmail.com}
\affiliation{Institute of Technical Physics, Department of Engineering Physics, Tsinghua University, Beijing 10084, P.R. China}
\author{Bai-Chu Yu}
\affiliation{Key Laboratory of Quantum Information, Chinese Academy of Sciences, School of Physics, University of Science and Technology of China, Hefei, Anhui, 230026, P.R. China}
\affiliation{Synergetic Innovation Center of Quantum Information and Quantum Physics, University of Science and Technology of China, Hefei, Anhui, 230026, P.R. China}
\author{Yu-Chun Wu}
\email{\tt wuyuchun@ustc.edu.cn}
\affiliation{Key Laboratory of Quantum Information, Chinese Academy of Sciences, School of Physics, University of Science and Technology of China, Hefei, Anhui, 230026, P.R. China}
\affiliation{Synergetic Innovation Center of Quantum Information and Quantum Physics, University of Science and Technology of China, Hefei, Anhui, 230026, P.R. China}
\author{Guang-Can Guo}
\affiliation{Key Laboratory of Quantum Information, Chinese Academy of Sciences, School of Physics, University of Science and Technology of China, Hefei, Anhui, 230026, P.R. China}
\affiliation{Synergetic Innovation Center of Quantum Information and Quantum Physics, University of Science and Technology of China, Hefei, Anhui, 230026, P.R. China}
\begin{abstract}
The celebrated Bell-Kochen-Specker no-go theorem asserts that quantum mechanics does not present the property of realism, the essence of the theorem is the lack of a joint probability distributions for some experiment settings. In this work, we exploit the information theoretic form of the theorem using information measure instead of probabilistic measure and indicate that quantum mechanics does not present such entropic realism neither. The entropic form of Gleason's no-disturbance principle is developed and it turns out to be characterized by the intersection of several entropic cones. Entropic contextuality and entropic nonlocality are investigated in depth in this framework. We show how one can construct monogamy relations using entropic cone and basic Shannon-type inequalities. The general criterion for several entropic tests to be monogamous is also developed, using the criterion, we demonstrate that entropic nonlocal correlations are monogamous, entropic contextuality tests are monogamous and entropic nonlocality and entropic contextuality are also monogamous. Finally, we analyze the entropic monogamy relations for multiparty and many-test case, which plays a crucial role in quantum network communication.
\end{abstract}
\maketitle

\emph{Introduction.}\textemdash It is common to assert that quantum mechanics (QM) is beyond the classical conception of nature. There are many phenomena which indicate that classical description of our universe is not suit for quantum world, for example, the uncertainty principle \cite{Heisenberg1927}, principle of complementarity \cite{bohr1928n,BANDYOPADHYAY2000233}, nonlocality \cite{bell1987einsteinpodolskyrosen,bell1966,Brunner2014bell},
contextuality \cite{kochen1967problem,Mermin1993}, negativity of quasi-probability \cite{Wigner1932} and so on. Among which contextuality and its more restricted form, nonlocality, attract many interests for both their theoretical importance and broad applications.
The works of Bell and  of Kochen and Specker indicated that quantum probabilities obtained from physical systems with dimension greater than two are incompatible with local hidden variable (LHV) model \cite{bell1964} and non-contextual hidden variable (NCHV) models \cite{kochen1967problem}, which is now known as Bell-Kochen-Specker (BKS) theorem. The essence of BKS theorem is that we can not preassign values to measurements in some consistent way, thus we can not always obtain a joint probability distributions for all involved measurements \cite{Fine1982}.

One of the most important properties of quantum probabilities is Gleason's  no-disturbance principle \cite{gleason1957measures,cabello2010non,Ramanathan2012,Jia2016},  which asserts that the outcome of a measurements $A$ does not depend on the (prior or simultaneous performed) compatible measurement $B$ and vice versa, QM is in accordance with the no-disturbance principle \cite{disturbance}.
When $A,B$ are implemented by two spatially separated parties, the no-disturbance principle is also known as nonsignalling principle, which characterizes the property that information can not be transformed faster than light. It has been shown that, in no-disturbance (nonsignalling) framework, contextuality (nonlocality) tests are monogamous \cite{Pawlowski2009monogamy,Ramanathan2012,Jia2016,jia2017exclusivity}, i.e., if one physicist observes the contextuality (nonlocality) in an experiment, then the other one must not observe the phenomenon simultaneously.

To understand these novel properties of quantum probabilities and find out the boundary between classical and quantum theory, we need to explore the notions from as many aspects as possible. Sheaf theoretic \cite{Abramsky2011,constantin2015sheaf} and graph theoretic \cite{Cabello2014graph,Yan2013,jia2017exclusivity} approaches have been well exploited. The information theoretic method provides a crucial and useful alternative, where the basic objects are information entropy of the probabilities arising in a given theory. Entropy is a crucial concept of thermodynamics, statistical mechanics and classical and quantum information theory \cite{wehrl1978general,Vedral2002}. It was Braunstein and Caves \cite{braunstein1988information} who first introduced the entropic form of Bell-Clauser-Horne-Shimony-Holt (CHSH) nonlocality tests \cite{bell1987einsteinpodolskyrosen,bell1966,CHSH}, then the entropic form \cite{Kurzynski2012} of Klyachko-Can-Binicio\u{g}lu-Shumovsky (KCBS) contextuality tests \cite{Klyachko2008simple} were also explored. Chaves, Fritz and Budroni \cite{Chaves2012,fritz2013entropic,Chaves2016} developed the approach in a more systematic way using the entropic cone and entropic inequalities \cite{yeung2008information} , they also developed the entropic description of nonsignalling correlations.

Despite of all these progresses, many works still need to be done. There is no existing information framework for Gleason's no-disturbance principle and the entropic form of Bell-Kochen-Specker theorem is less explored. The relationship between entropic tests and usual tests of contextuality and nonlocality is not very clear neither. Moreover, the monogamy relation, which is a characteristic feature of nonlocality and contextuality remains largely unexplored. In this work, we develop the information framework of Gleason's no-disturbance principle, the concepts of entropic nonlocality and entropic contextuality are introduced and their properties are extensively investigated  in this framework, and we present the entropic form of Bell-Kochen-Specker theorem. Using the tools of entropic vectors, entropic cone and computational geometrical method like Fourier-Motzkin elimination, we demonstrate the monogamy relations of entropic nonlocality and entropic contextuality. Finally, we explore the method to construct new monogamy relation from existing ones, and we extend the entropic monogamy relations to multiparty and many-test scenario, which will play an crucial role in realization of quantum network communication.

\emph{Information measure of quantum probabilities.}\textemdash
In generalized probability theory (which contains QM as a special case), a measurement can be regarded as some apparatus which accepts an input system and output random variables. If two measurements $A$ and $B$ are compatible, then they admit a joint probability distribution $p(a,b|A,B)$ where $a,b$ are random outcomes of $A,B$ respectively, and $p(a|A), p(b|B)$ can be reproduced from $p(a,b|A,B)$ by taking marginal distributions. To know how much information we can obtain from the physical system when we are measuring quantitatively, we need to introduce the concept of information measures which are various types of entropies and its siblings \cite{yeung2008information}. Here we use the terminology information measures to refer to entropy, conditional entropy, mutual information and conditional mutual information. The importance of information measures not only reflects in the widespread use in classical and quantum information theory \cite{yeung2008information,Nielsen2010,wilde_2017}, but also in statistical mechanics \cite{Jaynes1957,Jaynes1957a}, additive combinatorics \cite{Tao2010}, biodiversity studies \cite{Krebs1989} and so on.

For a given set of pairwise compatible measurements $\mathcal{M}$, there exists a joint entropy $H(\mathcal{M})$, and for any subset $\mathcal{S}\subseteq \mathcal{M}$, there also exists a joint entropy $H(\mathcal{S})$ which can be reproduced from $H(\mathcal{M})$. We now in a position to introduce the concepts of entropic vectors and entropic cones \cite{yeung2008information,Chaves2012,Chaves2016}. Let $2^{\mathcal{M}}$ be the power set of $\mathcal{M}$, the entropic vector is of the form $\mathbf{h}=[H(\mathcal{S}]_{\mathcal{S}\in 2^{\mathcal{M}}}$ where we make the convention that $H(\emptyset)=0$. For a given set of measurements, since (i) $\mathbf{0}$ is entropic; (ii) $\mathbf{h},\mathbf{h}'$ are entropic implies that $p\mathbf{h}+(1-p)\mathbf{h}'$ and $\alpha \mathbf{h}$ are entropic \cite{yeung2008information}, all entropic vectors form a cone named entropic cone, denoted as $\Gamma_{E}$.
Since every closed convex cone can be characterized by a system of linear inequalities, we may want to seek the precise linear inequalities which bound it. Notwithstanding, to give an explicit characterization of $\Gamma_E$ is still an open problem, an outer approximation is known, which is Shannon entropic cone $\Gamma_{SE}$ and is characterized by  Shannon-type inequalities \cite{Yeung1997}, clearly speaking, the nonnegative linear combinations of the following inequalities: (i) monotonicity $H(A|B)=H(AB)-H(B)\geq 0$;  (ii) subadditivity $I(A:B)=H(A)+H(B)-H(AB)\geq 0$; (iii) strong subadditivity ( or submodularity) $I(A:B|C)=H(AC)+H(BC)-H(ABC)-H(C)\geq 0$, where we use notation $A,B,C$ to represents some special measurements or collection of compatible measurements. It is worth mentioning that Shannon-type inequalities are not sufficient to completely characterize $\Gamma_E$ for $|\mathcal{M}|\geq 4$ \cite{yeung2008information}, they only give an outside approximation $\Gamma_{SE}$ , but they are the most explored and most useful ones and have been applied in many separated areas, including group theory \cite{yeung2008information}, Kolmogorov complexity \cite{HAMMER2000442}.

\emph{Entropic Bell-Kochen-Specker theorem.}\textemdash Given a set of measurements $\mathcal{M}=\{A_1,\cdots,A_n\}$, from QM we know that only compatible measurements are jointly measurable. We introduce the notion of compatible graph \cite{Jia2016} $G(\mathcal{M})$ for $\mathcal{M}$, where each measurement is represented as a vertex, and if two measurements are compatible, the corresponding vertices are connected by an edge. The Bell-Kochen-Specker (BKS) scenario \cite{bell1987einsteinpodolskyrosen,bell1966,kochen1967problem,Brunner2014bell,Chaves2012} $\mathcal{C}(\mathcal{M})$ for $\mathcal{M}$ is defined as $\mathcal{C}(\mathcal{M})=\{\mathcal{C}_1,\cdots,\mathcal{C}_k\}$ where $\mathcal{C}_i\subset \mathcal{M}$ and all measurements $A_{i_1},\cdots,A_{i_m}\in \mathcal{C}_i$ are pairwise compatible (they form a complete subgraph of $G(\mathcal{M})$), joint probability distributions $p(\mathcal{C}_i)=p(a_{i_1},\cdots,a_{i_m}|A_{i_1},\cdots,A_{i_m})$ are experimentally accessible, and for any $\mathcal{C}_i\subset \mathcal{C}_j$, $p(\mathcal{C}_i)$ can be reproduced from $p(\mathcal{C}_j)$ by taking marginal distribution. Collecting these probabilities together, we will have a vector $\mathbf{p}_{\mathrm{obs}}=[p(\mathcal{C}_k)]_{\mathcal{C}_k\in \mathcal{C}(\mathcal{M})}$ which we refer to as the behavior of $\mathcal{M}$ following the terminology introduced by Tsirelson \cite{tsirelson1993bs,Brunner2014bell}. If there exists a joint probability distribution $p(\mathcal{M})=p(a_1,\cdots,a_n|A_1,\cdots,A_n)$ for all measurements in $\mathcal{M}$ from which all experimentally accessible probabilities can be reproduced as marginal distributions, then we get a $2^{n}$-dimensional probabilistic vector $\mathbf{p}=[p(\emptyset),p(\mathcal{S}_1),\cdots,p(\mathcal{M})]$ \cite{dimension} where $\mathcal{S}_k$ are subsets of $\mathcal{M}$ and $p(\mathcal{S}_k)$ are marginal distributions of $p(\mathcal{M})$ \cite{pitowsky1989quantum,pitowsky1991pitowsky,pempty}. For a BKS scenario with observed probabilistic vector $\pobs$,
we know from Fine's theorem \cite{Fine1982} that the existence of NCHV/LHV model for a $\pobs$ is equivalent to the existence of a joint probability distribution for all involved measurements.
Thus observed data $\pobs$ is called non-contextual if it is consistent with a single well defined probability distribution $p(A_1,\cdots,A_n)$ for all measurements.
This further implies the existence of a probabilistic vector $\mathbf{p}$ which can project to $\pobs$. This is the well-known probabilistic vector formalism of contextuality and nonlocality.

Now, let us see how to understand contextuality and nonlocality using information measures of quantum probabilities. Suppose we are implementing measurements $\mathcal{M}$ with context set $\mathcal{C}(\mathcal{M})$, then the joint entropy $H(\mathcal{C}_k)$ for each $\mathcal{C}_k\in \mathcal{C}$ is experimentally accessible. Collecting all these data together we get an experimentally accessible entropic vector $\hobs=[H(\mathcal{C}_k)]_{\mathcal{C}_k\in \mathcal{C}(\mathcal{M})}$. The existence of joint probability distribution $p(\mathcal{M})$ imposes a strict constraints, contextual inequalities, on the experimentally accessible distributions $\pobs$. We can define entropic contextuality in a similar sense, $\hobs=[H(\mathcal{C}_k)]_{\mathcal{C}_k\in \mathcal{C}(\mathcal{M})}$ is called entropic non-contextual or entropic nonlocal if all entropies  $H_{\mathcal{C}_k}$ are consistent with a single well defined shannon entropy $H_{\mathcal{M}}$, i.e., there exist an entropic vector $\mathbf{h}=\{H(S_k)\}_{\mathcal{S}_k\in 2^{\mathcal{M}}}$ which can project to $\hobs$. As have been proved for single system \cite{Kurzynski2012,Chaves2012} and for composite system \cite{braunstein1988information},  QM is inconsistent with the entropic NCHV/LNV models, we refer to such kind of contextuality and nonlocality as entropic ones. Therefore we obtain the following entropic Bell-Kochen-Specker theorem:
\begin{thm}[Entropic Bell-Kochen-Specker]For physical systems of dimension greater than two, there is no consistent way to assign an corresponding entropic vectors $\mathbf{h}(\mathcal{M})$ which can project to the observed entropic vectors $\hobs(\mathcal{M})$ for all set of measurements $\mathcal{M}$.
\end{thm}

\begin{figure}
  \centering
  \includegraphics[width=5.5cm]{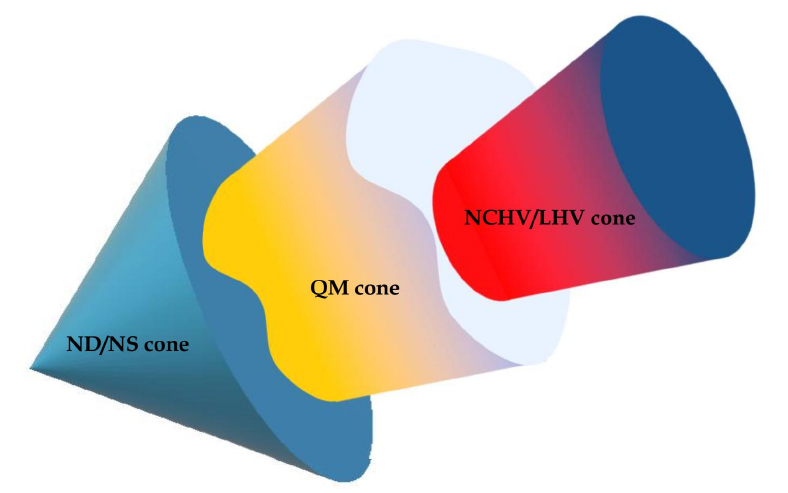}\\
  \caption{Pictorial illustration of the  non-disturbance (ND) or nonsignaling (NS) cone, quantum mechanical (QM) cone and non-contextual hidden variable (NCHV) or local hidden variable (LHV) cone. }\label{fig:cone}
\end{figure}

We are now in a position to give some criteria to detect the entropic contextuality. Suppose that we have a set of $n$ measurements $\mathcal{M}=\{A_1,\cdots,A_n\}$ and the BKS scenario is
$\mathcal{C}(\mathcal{M})=\{\{A_1,A_2\},\{A_2,A_3\},\cdots,\{A_n,A_1\}\}$, i.e., each $A_i$ and $A_{i+1}$ pair is compatible, thus there is a joint entropy $H(A_i,A_{i+1})$. If $\hobs$ is entropic non-contextual, then there exist a Shannon entropy $H(A_1\cdots,A_n)$ which is consistent whit all $H(A_i,A_{i+1})$. Repeatedly using the chain rule $H(A_i,A_i+1)= H(A_i|A_{i+1})+H(A_{i+1})$ and monotonicity conditions $H(A_i|A_{i+1})\leq H(A_i)\leq H(A_i,A_i+1)$ which is nothing more than the Shannon-type inequalities, we obtain the inequality
 \begin{equation}\label{eq:e-contextuality}
\mathcal{I}_{\mathcal{M}}^{cycle}= H(A_1|A_n)-\sum_{i=2}^nH(A_1|A_i)\leq 0.
 \end{equation}
It has been shown theoretically \cite{Kurzynski2012,Chaves2012} and then verified experimentally \cite{Zhan2017} in three level system with five measurements, the inequality can be violated by QM.
For composite system, distant Alice and Bob choose to implement $2m$ measurements $A_0,\cdots,A_{m-1}$ and $B_0,\cdots,B_{m-1}$ with BKS scenario $\mathcal{C}(\mathcal{M})=\{\{A_i,B_j\}_{i,j=0}^{m}\}$, then the entropic test inequalities reads
\begin{align}
\mathcal{B}^{m}=& H(A_0|B_{m-1})-[H(A_0|B_0)+ H(B_0|A_1)+ \cdots\nonumber\\
& +H(A_{m-1}|B_{m-1})]\leq 0.
\end{align}
The compatible gragh of above entropic Bell inequality is a $2m$ bipartite cycle graph. It's show that QM can violate the inequalities for $m=2$ case \cite{braunstein1988information}, which we also refer to as entropic CHSH test, and denoted as $\mathcal{B}^{CHSH}$. Entropic vectors $\mathbf{h}$ which admits an entropic NCHV/LHV model will form a cone named NCHV/LHV cone, denoted as $\Gamma_{NCHV}$ or $\Gamma_{LHV}$, see Fig. \ref{fig:cone} for a pictorial illustration. Note that since there exist inequalities which are not of Shannon-type, NCHV/LHV entropic cone based on Shannon-type inequalities are bigger than the true NCHV/LHV entropic cone, that is there exist some observed entropic vectors $\hobs$ which exhibit entropic contextuality but can not be detected in this way. Nevertheless, entropic cone obtained in this way is a good and useful approximation.

\emph{Entropic no-disturbance principle.}\textemdash No-disturbance principle asserts that compatible measurements do not disturb the experimentally accessible statistics of each other, it is a physically motivated principle proposed to explain the weirdness of QM \cite{Ramanathan2012,Jia2016}. Mathematically, it reads $p(a|A)=\sum_bp(a,b|A,B)=\sum_{b'}p(a,b'|A,B')$ for any $B,B'$ compatible with $A$. Geometrically, the no-disturbance principle is characterized by the intersection of several probabilistic simplex polytopes. For instance, to characterize no-disturbance principle for KCBS test, the experimentally accessible data is $\pobs=\{p(a_1,a_2|A_1,A_2),\cdots,p(a_5,a_1|A_5,A_1)\}$ which is completely characterized by $5\times 2^2$ probabilities (since $A_1,\cdots,A_5$ are all dichotomic measurements) and thus $\pobs \in \mathbb{R}^{5\times 2^2}$. There are two constraints imposed by properties of probabilities: (i) positivity $\pobs \in \mathbb{R}_{\geq 0}^{5\times 2^2}$; and (ii) normalization $\sum_{a_i,a_{i+1}}p(a_ia_{i+1}|A_iA_{i+1})=1$. The no-disturbance constraints are then set of equalities: $\sum_{a_{i-1}}p(a_{i-1}a_{i}|A_{i-1}A_{i})=\sum_{a_{i+1}}p(a_ia_{i+1}|A_iA_{i+1})$, thus no-disturbance vector form a affine subspace of $\mathbb{R}_{\geq 0}^{5\times 2^2}$ which is the intersection of subspace $\mathbb{P}_{i}$ restricted by $\sum_{a_{i-1}}p(a_{i-1}a_{i}|A_{i-1}A_{i})=\sum_{a_{i+1}}p(a_ia_{i+1}|A_iA_{i+1})$, i.e., $\mathbb{P}_{ND}=\mathbb{P}_{1}\cap\cdots \cap \mathbb{P}_{5}$.

In entropic formalism, to formalize the fact that the outcomes of a measurement do not depend on the outcomes of the measurements  compatible with it, we need to introduce the notion of no-disturbance cone. For a given BKS scenario $[\mathcal{M},\mathcal{C}(\mathcal{M})]$ with $\mathcal{C}(\mathcal{M})=\{\mathcal{C}_1,\cdots,\mathcal{C}_m\}$, there are $m$ corresponding entropic cones $\Gamma_{\mathcal{C}_k}$ which are characterized by the basic inequalities for measurements contained in $\mathcal{C}_k$. The no-disturbance cone is defined as the intersection of all entropic cones corresponding to $\mathcal{C}_k\in \mathcal{C}$, i.e., $\Gamma_{ND}=\Gamma_{\mathcal{C}_1}\cap\cdots\cap \Gamma_{\mathcal{C}_m}$. This can be understand as follows: all $\Gamma_{\mathcal{C}_k}$ are characterized by the corresponding basic inequalities, when embedding the entropic vector in $\Gamma_{\mathcal{C}_k}$ into a bigger space, the vectors not contained in it are not constrained. But if the vectors $\hobs$ contained in both $\Gamma_{\mathcal{C}_k}$ and $\Gamma_{\mathcal{C}_l}$, it must satisfy the inequalities imposed by two cones simultaneously. This is consistent with the probabilistic formalism. For a given BKS scenario, as depicted in Fig. \ref{fig:cone}, there is a clear hierarchy among the NCHV/LHV cone $\Gamma_{NCHV/LHV}$, QM cone $\Gamma_{QM}$ and no-disturbance cone $\Gamma_{ND}$: $\Gamma_{NCHV/LHV}\subset \Gamma_{QM}\subset \Gamma_{ND}$.

\emph{Monogamy relations in entropic no-disturbance framework.}\textemdash
Quantum contextuality and nonlocality are captured by the violations of the corresponding test inequalities, one of the characteristic features of these inequalities is monogamy relations \cite{Pawlowski2009monogamy,Ramanathan2012,Jia2016,jia2017exclusivity} which play crucial roles in quantum key distribution \cite{Barrett2005,Pawlowski2010security} and quantum network \cite{Lee}. The monogamy relation says that if one physicist observes the contextuality or nonlocality in some experiment, then all other simultaneous test experiments (with measurement settings having some compatible part) can not observe the violated values. It is natural to ask which part of QM is responsible for monogamy, no-disturbance (nonsignalling) principle is regarded as a hopeful candidate \cite{Pawlowski2009monogamy,Ramanathan2012,Jia2016}. One may ask if entropic contextuality and nonlocality also present monogamy relations and if entropic no-disturbance can be used to explain such kind of monogamy. Actually, this is the case, we are going to show that, in entropic no-disturbance framework entropic contextuality tests, entropic nonlocality tests, and entropic contextuality and nonlocality tests all present monogamy.

Let us first illustrate how to construct a joint entropic vector of two given entropic vectors which obey entropic no-disturbance principle. For vectors $\mathbf{h}(XY)=[H(\emptyset), H(X), H(Y), H(XY)]$ and $\mathbf{h}(XZ)=[H(\emptyset), H(X), H(Z), H(XZ)]$, we can construct $H(XYZ)=H(XY)+H(XZ)-H(X)$ and thus a joint entropic vector $\mathbf{h}(XYZ)$ which can project to $\mathbf{h}(XY)$ and $\mathbf{h}(XZ)$. Using this strategy, we construct joint entropic vector for any set of measurements with chordal compatible graph.
\begin{lemma}
For a given set of measurements $\mathcal{M}=\{A_1,\cdots,A_n\}$, if the compatible graph $G(\mathcal{M})$ is a chordal graph for which there exists no induced cycle of more than $3$ edges, then for each experimentally accessible entropic vectors $\hobs(\mathcal{M})$  which obey the entropic no-disturbance principle, there exists an entropic vector $\mathbf{h}(\mathcal{M})$ which can project to $\hobs(\mathcal{M})$.
\end{lemma}

The proof is similar as in \cite{Ramanathan2012}, see supplemental material \cite{supp} for details. Note that the construction is a direct derivation of the entropic no-disturbance, we are now to investigate the monogamy relations of contextuality and nonlocality tests. First note that the existence of a joint entropic vector $\mathbf{h}(\mathcal{M})$ for all $\hobs(\mathcal{M})$ which we obtain from a QM experiment will lead to the unviolated value of the test inequality $\mathcal{I}_{\mathcal{M}}\overset{Q}{\leq}0$, which further means that in QM framework the experiment can not reveal the contextuality or nonlocality. We then give the following monogamy criterion.
\begin{thm}[Entropic monogamy criterion]\label{thm:monogamy}
Suppose two couple of physicists are simultaneously running two contextuality or nonlocality tests $\mathcal{E}_1$ and $\mathcal{E}_2$ with measurement sets $\mathcal{M}_1$ and $\mathcal{M}_2$ and test inequalities $\mathcal{I}_{\mathcal{M}_1}\leq 0$ and $\mathcal{I}_{\mathcal{M}_2}\leq 0$ respectively. If the compatible graph $G(\mathcal{M}_1\cup \mathcal{M}_2)$ corresponding to all involved measurements can be decomposed into two chordal graphs, then two tests are monogamous:
\begin{equation}\label{}
  \mathcal{I}_{\mathcal{M}_1}+\mathcal{I}_{\mathcal{M}_2}\overset{Q}{\leq} 0.
\end{equation}
\end{thm}

See supplemental material \cite{supp} for the detailed proof. To make things more concise, let us see some examples. First we note that the monogamy of entropic CHSH inequalities has be proved by Chaves and Budroni \cite{Chaves2016} in the entropic nonsignalling framework (which is the special case of entropic no-disturbance principle), $\mathcal{B}^{CHSH}_{\mathcal{\mathcal{M}}}+\mathcal{B}^{CHSH}_{\mathcal{M'}}\overset{Q}{\leq} 0$. With the similar strategy, we can also prove the monogamy of contextuality
\begin{equation}\label{eq:mono-NL-C}
\mathcal{I}^{cycle}_{\mathcal{\mathcal{M}}}+\mathcal{I}^{cycle}_{\mathcal{M'}}\overset{Q}{\leq} 0,
\end{equation}
where $\mathcal{M}=\{A_1,\cdots,A_n\}$, $\mathcal{M}=\{A'_1,\cdots,A'_m\}$, $A_2=A'_2$, $A_n=A'_m$, $A_1$ is compatible with all $A'_3,\cdots,A'_{m-1}$ and $A'_1$ is compatible with all $A_3,\cdots,A_{n-1}$; And monogamy of entropic nonlocality and entropic contextuality
 \begin{equation}\label{eq:mono-NL-C}
\mathcal{B}^{CHSH}+\mathcal{I}^{KCBS}\overset{Q}{\leq} 0.
\end{equation}
See supplementary material \cite{supp} for detailed proof of all above monogamy relations.

\emph{Constructing new monogamy relations from old ones.}\textemdash
Here we analyze how to construct new monogamy relations from the old ones, we will take entropic CHSH inequality as a prototypical example and the generalization to other cases are straightforward. Actually, as what have been done patially in~\cite{Jia2016}, if we have some monogamy relations in hand, we can construct more monogamy relations. For example, from $\mathcal{B}^{CHSH}_{ij}+\mathcal{S}^{CHSH}_{jk}\overset{Q}{\leq} 0$ where $i,j,k$ label different parties, we can construct a new cycle monogamy relation,
\begin{equation}\label{eq:star-monogamy}
\sum_{i=0}^{m-1}\mathcal{B}^{CHSH}_{ij}\overset{Q}{\leq} 0,
\end{equation}
where we assume that all overlapped parties share their measurements in each entropic CHSH experiment. To see how this holds, we only need to copy each experiments, i.e., we have $2 \sum_{i=1}^{m}\mathcal{B}^{CHSH}_{ij}$, then we can divide them into $m$ pairs of two-monogamy sums, e.g., $\mathcal{B}^{CHSH}_{ij}+\mathcal{B}^{CHSH}_{i+1j}$ and sum in the subscript is module $m$, then we arrive at the Eq. (\ref{eq:star-monogamy}). With this method we obtain:
\begin{corollary}\label{cor:mono}
For $n$ parties labeled as $\mathcal{P}=\{1,\cdots, n\}$ (drawn as vertices of the party graph $G(\mathcal{P})$) who implementing several entropic CHSH experiments $\mathcal{B}^{CHSH}_{ij}$ (represented as edges of the party graph $G(\mathcal{P})$) simultaneously and all measurements of the overlapped parties are shared in each experiment, if the party graph $G(\mathcal{P})$ is a connected graph which can be packed into some small pieces of Euler graphs, complete graphs, $2k$-edge lines, stars  and cycles, then these experiments are monogamous,
\begin{equation}\label{eq:n-monogamy}
  \sum_{(ij)\in E[G(\mathcal{P})]}\mathcal{B}^{CHSH}_{ij} \overset{Q}{\leq} 0.
\end{equation}
See supplemental material \cite{supp} for detailed proof.
\end{corollary}
Notice that for other type of entropic contextuality and nonlocality tests, we have the similar result. Monogamy relation plays an important role in quantum key distribution processes \cite{Barrett2005,Pawlowski2010security}. But most results concentrated on the two-test case, this corollary provides a generalization to multiparty and many-tests case, which will be useful for quantum network \cite{Lee}. For theories of which entropic no-disturbance principle does not hold, it is possible to observe violated values of two BKS tests simultaneously. However, the causal structure is broken in this case as argued probabilistic case in \cite{Ramanathan2012}. More specifically, if two entropic nonsignalling Bell tests are not monogamous in some theory, then there must be superluminal communications among three parties, this would broke the causal structure; if two entropic no-disturbance Kochen-Specker test monogamy are violated, the entropic vectors obtained from the first experiment settings and from the compatible part of the other experiment are different, then we can using the difference to propagating influence backward in time if two experiments are implemented sequentially, which would also lead a violation of causality.

\emph{Conclusions and discussions.}\textemdash
Entropic formalism is an  alternative approach to understand quantum contextuality and nonlocality \cite{Chaves2012,fritz2013entropic,Chaves2016}, it has several advantages: this formalism does not depend on the outcomes of the measurement as strongly as usual probabilistic formalism thus they can be applied to systems with arbitrary dimensions, and the detection inefficiencies is well dealt in this formalism. We exploited the entropic form of Bell-Kochen-Specker no-go theorem. Then we demonstrated that monogamy of Bell-Kochen-Specker tests seen in probabilistic formalism can be extended to the entropic formalism. Adopting the graph theoretic method, we develop the general criterion of monogamy of several Bell-Kochen-Specker tests. And we all analyze the multiparty and many-test case, which will be crucial for quantum network communication.

\begin{acknowledgments}
This work was supported by the National Key Research and Development Program
of China (Grant No. 2016YFA0301700) and the Anhui Initiative in Quantum Information Technologies
(Grants No. AHY080000)
\end{acknowledgments}


\vspace{\columnsep}

\appendix



\section*{Supplemental Material}

\section{Entropic vector formalism}

\subsection{Probabilistic vectors and observed probabilistic vectors}
In a typical contextuality test experiment, we deal with the observed probabilities involved in the experiment. For example,
for KCBS test experiment \cite{Klyachko2008simple}, the experimentally accessible probabilities are
$\mathbf{p}_{\mathrm{obs}}$ $=[p(a_1a_2|A_1A_2)$, $p(a_2a_3|A_2A_3)$, $p(a_3a_4|A_3A_4)$, $p(a_4a_5|A_4A_5)$, $p(a_5a_1|A_5A_1)]$. After we obtain the probabilities, we need to calculate a test parameter
\begin{equation}\label{eq:KCBS}
\sum_{i=1}^{5}\langle A_iA_{i+1}\rangle \geq -3,
\end{equation}
if the inequality is violated, then the observed probabilities are said to be contextual.
For the general case, where we have a set of measurements $\mathcal{M}=\{A_1,\cdots,A_n\}$, some of which are compatible, thus the corresponding joint probabilities are experimentally accessible. A probabilistic vector is
$\mathbf{p}=[p(\emptyset),p(\mathcal{S}_1),\cdots,p(\mathcal{M})]$ where $\mathcal{S}_i$ is a subset of $\mathcal{M}$ and all subsets are contained.
We also make the convention that $p(\emptyset)=0$. The observed probabilistic vector is defined as the probabilistic vectors which only contain the experimentally accessible probabilities, denoted as $\pobs$.
We can project a probabilistic vector into an observed probabilistic vector. An observed probabilistic vector $\pobs$ is said to be non-contextual if there exists a probabilistic vector $\mathbf{p}$ which can be projected to $\pobs$.

Let us take $n$-cycle contextuality test scenario \cite{Araujo2013n-cycle} as an example to illustrate all involved notions: a set of $n$ measurements $\mathcal{M}_n=\{A_1,\cdots,A_n\}$ is chosen to implemented, and each $A_i$ is compatible with $A_{i+1}$ (we make the convention that $n+1=1$), thus the corresponding compatible graph is an $n$-cycle. The corresponding BKS scenario is $\mathcal{C}_n=\{\{A_1,A_2\},\cdots,\{A_n,A_1\}\}$, and the corresponding behavior is $\mathbf{p}_{\mathrm{obs}}=[p(a_1,a_2|A_1,A_2),\cdots, p(a_n,a_1|A_n,A_1)]$. To test if $\mathbf{p}_{\mathrm{obs}}$ admits an NCHV model, we need to check a test parameter $\mathcal{I}=\sum_{i=1}^n \gamma_i \langle A_{i}A_{i+1}\rangle$ where $\gamma_i=\pm 1$ and the number of minus ones is odd, and find out if $\mathcal{I}\leq n-2$ or not. If we choose all measurements be dichotomic with outcomes $\pm 1$, then, the test parameter can also be written as
$\mathcal{I}'=\sum_{a\oplus b=0}\sum_{\begin{subarray}{lcl}
          i,\gamma_i=1\end{subarray}} p(ab|ii+1)+\sum_{a\oplus b\neq 0}\sum_{\begin{subarray}{lcl}
          i,\gamma_i=-1\end{subarray}} p(ab|ii+1)$ and the corresponding inequality is $\mathcal{I}'\overset{NCHV}{\leq}n-1$.
It is clear that all above inequalities are just the inner product $\langle\mathbf{s},\mathbf{p}_{\mathrm{obs}}\rangle$ for $\mathbf{s}$ are obtained from test parameter, thus they are all linear constraints on the behavior we have access to. The probabilistic vectors allowed by NCHV model then form a polytope known as NCHV polytope.

\begin{figure}
  \centering
  \includegraphics[scale=0.4]{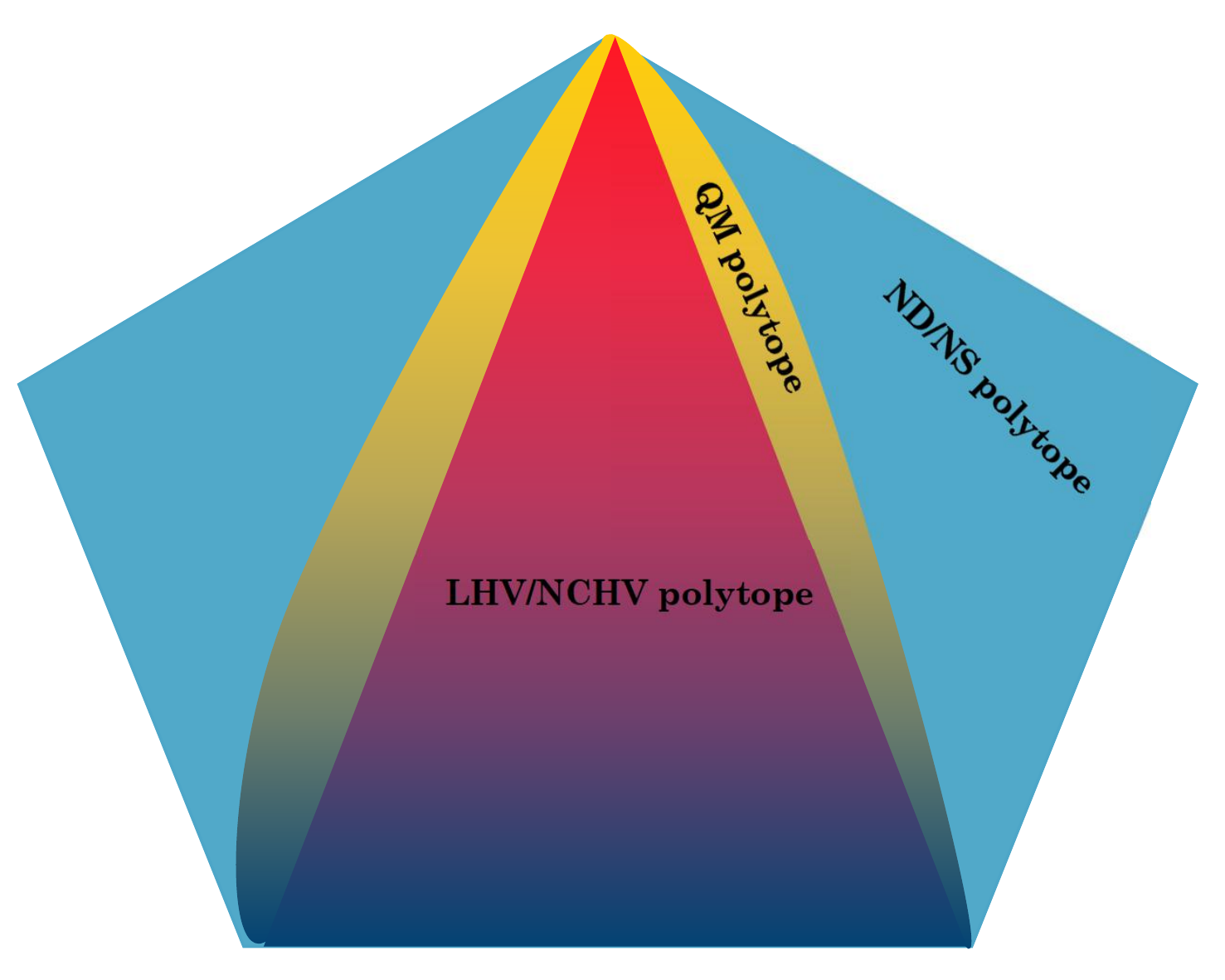}\\
  \caption{Pictorial illustration of the  non-disturbance (ND) or nonsignaling (NS) polytope, quantum mechanical (QM) polytope and non-contextual hidden variable (NCHV) or local hidden variable (LHV) polytope. }\label{fig:polytope}
\end{figure}

No-disturbance principle asserts that compatible measurements do not disturb the experimentally accessible statistics of each other, mathematically, it reads $p(a|A)=\sum_bp(a,b|A,B)=\sum_{b'}p(a,b'|A,B')$ for any $B,B'$ compatible with $A$. Geometrically, the no-disturbance principle is characterized by the intersection of several probabilistic simplex polytopes. See Fig. \ref{fig:polytope} for the pictorial illustration of the  non-disturbance (ND) or nonsignaling (NS) polytope, quantum mechanical (QM) polytope and non-contextual hidden variable (NCHV) or local hidden variable (LHV) polytope.
\subsection{Entropic vector and Entropic cone}

Let us take a close look at the entropic cone here, see Ref. \cite{yeung2008information} for more details. we use the notation $\Nb_n$ to mean the set of first $n$ natural numbers, $\{1,2,\cdots,n\}$, which we utilize as a label set to label $n$ random variables
\[\mathcal{M}=\{X_i|i\in \Nb_n\}.\]
The power set of $\Nb_n$ is denoted as $2^{\Nb_n}$ whose number of elements is $2^n$. Associated with $\mathcal{M}$ are $2^{n}-1$ joint entropies. For instance, for $\Nb_2$, we have $\{H(X_1),H(X_2),H(X_1,X_2)\}$.

Now let $\alpha$ be a subset of $\Nb_n$, i.e., $\alpha\in 2^{\Nb_n}$, and let
$C_{\alpha}=\{X_i|i\in \alpha\}$ which we refer to as a context of $\mathcal{M}$ and correspondingly $H_{\mathcal{M}}(C_{\alpha})=H(X_\alpha\}$. We will make the convention that $H(\emptyset)=0$. Consider the $2^n$ dimensional real vector space $\mathbb{R}^{2^n}$, a vector $\mathbf{h}=(h_\alpha)_{\alpha \in 2^{\Nb_n}}$ is called a entropic vector if it is the vectors of joint entropies of a set of $n$ random variables $\mathcal{M}$, i.e., $h_{\alpha}=H(X_{\alpha})$ for all $\alpha\in 2^{\Nb_n}$. Notice that $H$ can be chosen as any information measure (any entropy function), not restricted to Shannon entropy.
We now analyze the region (which we refer to as entropic quasi-cone)
\begin{equation}\label{eq:region}
\Gamma^{*}_{E}=\{\mathbf{h}|\mathbf{h}\mathrm{\,is\, entropic}\}.
\end{equation}

We first list some important properties of $\Gamma_{E}^{*}$:
\begin{itemize}
  \item[(1)] $\Gamma_{E}^{*}$ contains the origin point of $\mathbb{R}^{2^n}$;
  \item[(2)] $\Gamma_{E}^{*}$ is a subset of the nonnegative orthant $\mathbb{R}_{\geq 0}^{2^n}$;
  \item[(3)] the closure $\Gamma_{E}^{*}$, denoted as $\bar{\Gamma}_{E}^{*}$,  is convex.
\end{itemize}
From the above properties we know that
\begin{equation}\label{eq:cone}
  \Gamma_{E}=\bar{\Gamma}_{E}^{*},
\end{equation}
is a convex cone, which we will refer to as entropic cone.

We also need to introduce the notion entropic expression, which is a linear function of the entropies $f(H(X),H(X|Y),\cdots)$. An entropic inequality is of the form
\begin{equation}\label{eq:e-inequality}
f(H(X),H(X|Y),\cdots)\geq c,
\end{equation}
or we can make it into $f(H(X),H(X|Y),\cdots)-c\geq 0$. The Shannon entropic cone is characterized by Shannon-type inequalities: (i) monotonicity $H(A|B)=H(AB)-H(B)\geq 0$;  (ii) subadditivity $I(A:B)=H(A)+H(B)-H(AB)\geq 0$; (iii) strong subadditivity ( or submodularity) $I(A:B|C)=H(AC)+H(BC)-H(ABC)-H(C)\geq 0$, where we use notation $A,B,C$ to represents some special measurements or collection of compatible measurements.

Like in probabilistic formalism, for a given set of measurements $\mathcal{M}$, we have access to the joint entropic $H(\mathcal{S})$ compatible measurements $A_i\in \mathcal{S}$, collecting these entropies together we will get the observed entropic vector $\hobs$. If there exist a joint entropy $H(\mathcal{M})$ for all involved measurements $A_i\in\mathcal{M}$ (thus an entropic vector $\mathbf{h}=[H(\mathcal{S})]_{\mathcal{S}\in 2^{\mathcal{M}}})$, for which all experimentally accessible entropies can be reproduced from this entropy, then we say than $\hobs$ is entropic noncontextual. All entropic noncontextual entropic forms a cone named as NCHV cone; entropic vectors arise in QM form a cone called QM cone; entropic vectors satisfy entropic no-disturbance principle form a cone we refer to as ND cone. There is a hierarchy structure of these cones, see Fig. \ref{fig:cone}.

\section{Proofs of monogamy relations}
In this section, we give the detailed proof the the theorem appeared in the main text.

\subsection{General monogamy criterion}
\begin{lemma*}
For a given set of measurements $\mathcal{M}=\{A_1,\cdots,A_n\}$, if the compatible graph $G(\mathcal{M})$ is a chordal graph for which there exists no induced cycle of more than $3$ edges, then for each experimentally accessible entropic vectors $\hobs(\mathcal{M})$  which obey the entropic no-disturbance principle, there exists an entropic vector $\mathbf{h}(\mathcal{M})$ which can project to $\hobs(\mathcal{M})$.
\begin{pf}
Inspired by the construction of Ramanathan \emph{et al.} \cite{Ramanathan2012}, we give a explicit entropic construction. Since $G(\mathcal{M})$ is a chordal graph, thus $G(\mathcal{M})$ contains no induced cycle of length greater than $3$. Let $V[G(\mathcal{M})]$ be the set of vertices of $G(\mathcal{M})$, $G_{3}(\mathcal{M})=\{K_3^i\}$ be the set of all complete subgraphs (all vertices are pairwise connected) with more than or equal to $3$ vertices, $G_{2}(\mathcal{M})=\{E_2^j\}$ be the set of edges which are not subgraphs of any $K_3^i\in G_{3}(\mathcal{M})$, $G_1(\mathcal{M})=\{V_1^k\}$ be the set of vertices which are not subgraphs of any $K_3^i\in G_{3}(\mathcal{M})$ and $E_2^j\in G_{2}(\mathcal{M})$ and $\mathcal{G}=G_{3}(\mathcal{M})\cup G_{2}(\mathcal{M}) \cup G_1(\mathcal{M})$. From the construction, it is clear that each edge of $G(\mathcal{M})$ contained only one of $G_3(\mathcal{M})$ or $G_2(\mathcal{M})$ but they can appear more than once in each set. Now we can construct a joint entropy $H(\mathcal{M})=H(A_1,\cdots,A_n)$ for $H(K_3^i)$, $H(E_2^j)$ and $H(V_1^k)$:
\begin{align}
  H(\mathcal{M}) &= \sum_{K_3^i\in G_{3}(\mathcal{M})}H(K_3^i)+\sum_{E_3^j\in G_{2}(\mathcal{M})}H(E_2^j)\nonumber \\
   & +\sum_{V_1^k\in G_1(\mathcal{M})}H(V_1^k)\nonumber\\
   &-\frac{1}{2}\sum_{K^l\neq K^s,K^l,K^s\in \mathcal{G}}H(K^l\cap K^s).
\end{align}
Thus we can construct the entropic vector $\mathbf{h}(\mathcal{M})$ which can project to $\hobs=[H(K)]$ where $K$ is taken over all complete subgraphs of $G(\mathcal{M})$, therefore we completes the proof.\qed
\end{pf}
\end{lemma*}

\begin{theorem*}[Entropic monogamy criterion]
Suppose two couple of physicists are simultaneously running two contextuality or nonlocality tests $\mathcal{E}_1$ and $\mathcal{E}_2$ with measurement sets $\mathcal{M}_1$ and $\mathcal{M}_2$ and test inequalities $\mathcal{I}_{\mathcal{M}_1}\leq 0$ and $\mathcal{I}_{\mathcal{M}_2}\leq 0$ respectively. If the compatible graph $G(\mathcal{M}_1\cup \mathcal{M}_2)$ corresponding to all involved measurements can be decomposed into two chordal graphs, then two tests are monogamous:
\begin{equation}\label{}
  \mathcal{I}_{\mathcal{M}_1}+\mathcal{I}_{\mathcal{M}_2}\overset{Q}{\leq} 0.
\end{equation}
\begin{pf}
Suppose $G(\mathcal{M}_1\cup \mathcal{M}_2)$ can be decomposed into two chordal graph $G(\mathcal{N}_1)$ and $G(\mathcal{N}_{2})$ where $\mathcal{N}_1\cup \mathcal{N}_2=\mathcal{M}_1\cup \mathcal{M}_2$, then there always exist two entropic vectors $\mathbf{h}(\mathcal{N}_1)$ and $\mathbf{h}(\mathcal{N}_{2})$ which can project to $\hobs(\mathcal{N}_1)$ and $\hobs(\mathcal{N}_2)$, thus the correspond tests of $\mathcal{N}_1$ and $\mathcal{N}_2$ can not get violated value in QM framework, i.e., $\mathcal{I}_{\mathcal{N}_1}\overset{Q}{\leq}0$ and $\mathcal{I}_{\mathcal{N}_2}\overset{Q}{\leq}0$. Since we have $\mathcal{I}_{\mathcal{M}_1}+\mathcal{I}_{\mathcal{M}_2}=\mathcal{I}_{\mathcal{N}_1}+\mathcal{I}_{\mathcal{N}_2}$, we get the required result.\qed
\end{pf}
\end{theorem*}

\begin{figure}
\includegraphics[scale=0.7]{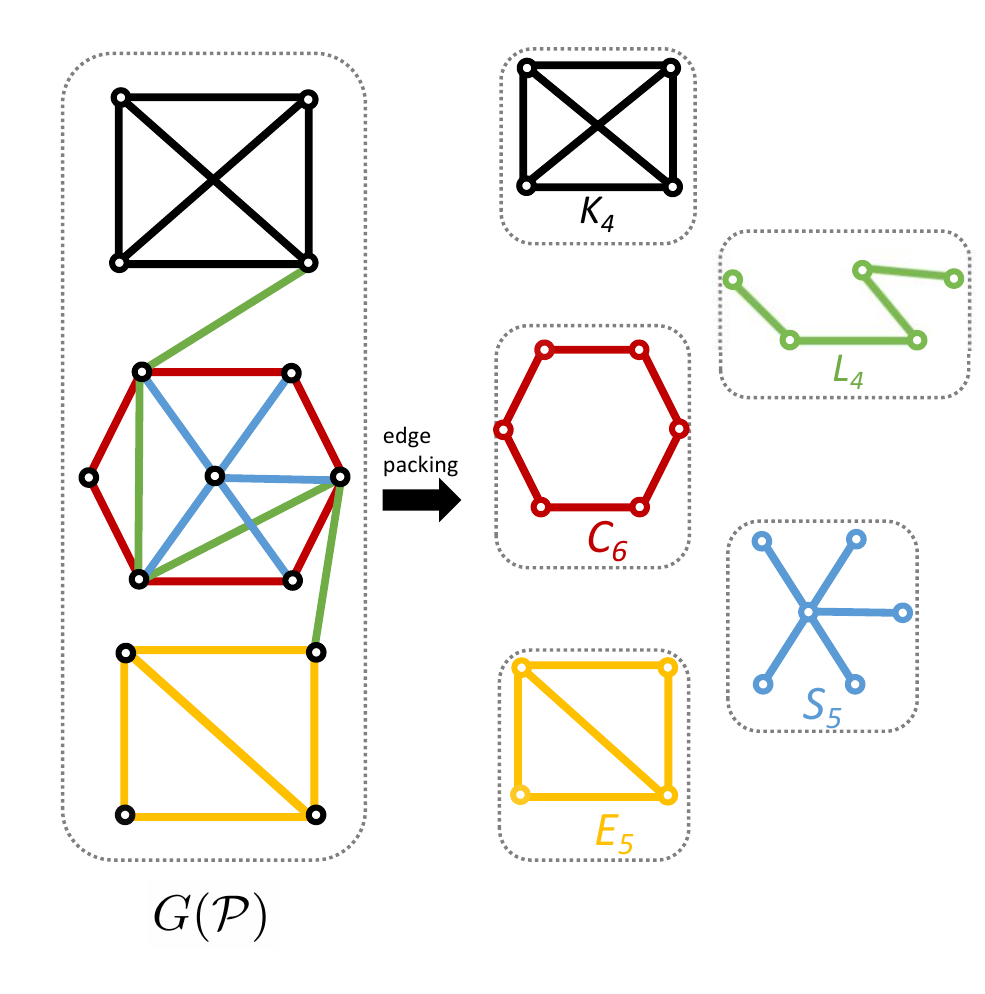}
\caption{\label{fig:general_monogamy}  The depiction of the monogamy relation of entropic CHSH experiments among many parties, each vertex represents a party and each edge represents a entropic CHSH experiment. On the left side, it is the whole party graph $G(\mathcal{P})$ of all involved experiments. The edge set of $G(\mathcal{P})$ can be packed into five small pieces as a 4-complete graph $K_4$, a 6-cycle graph $C_6$, a 5-Euler graph $E_5$, a 4-edge line $L_4$ and a 5-star graph $S_5$. Since all these small pieces are monogamous then using the triangle inequality of absolute value function, we can get the monogamy relation corresponding to $G(\mathcal{P})$.}
\end{figure}

\begin{corollary*}
For $n$ parties which are labeled as $\mathcal{P}=\{0,\cdots, n\}$ (represented as vertices of the party graph $G(\mathcal{P})$) who implementing several entropic CHSH experiments $\mathcal{B}^{CHSH}_{ij}$ (represented as edges of the party graph $G(\mathcal{P})$) simultaneously and all measurements of the overlapped parties are shared in each experiment, if the party graph $G(\mathcal{P})$ is a connected graph which can be divided into some small pieces of Euler graphs, complete graphs, $2k$-edge lines, stars  and cycles, then these experiments are monogamous, i.e.
\begin{equation}
  \sum_{(ij)\in E[G(\mathcal{P})]}\mathcal{B}^{CHSH}_{ij} \overset{Q}{\leq} 0.
\end{equation}
\begin{pf}
The process is actually the edge packing, since we can using the copy method to make the sum of all individual edges into the sum of the groups of edges of each packed small pieces, we only need to show that there is a monogamy relation for each small piece, i.e., we need to show that there is monogamy relations for Euler graphs, complete graphs, $2k$-edge lines, stars  and cycles, this can be done in the same spirit as we have done for star graph case for Eq. (\ref{eq:star-monogamy}). For example, as depicted in Fig. \ref{fig:general_monogamy}, we have the whole party graph $G(\mathcal{P})$, since its edge set $E[G(\mathcal{P})]$ can be packed into five small pieces, a 4-complete graph $K_4$, 6-cycle graph $C_6$, a 5-Euler graph $E_5$, a 4-edge line $L_4$ and a 5-star graph $S_5$ then we have
\begin{align}\label{}
& \sum_{(ij)\in E[G(\mathcal{P})]}\mathcal{B}^{CHSH}_{ij}\nonumber\\
=& \sum_{(ij)\in K_4}\mathcal{B}^{CHSH}_{ij}+ \sum_{(ij)\in C_6} \mathcal{B}^{CHSH}_{ij} +\sum_{(ij)\in S_5} \mathcal{B}^{CHSH}_{ij}\nonumber \\
+& \sum_{(ij)\in E_5} \mathcal{B}^{CHSH}_{ij} + \sum_{(ij)\in L_4}\mathcal{B}^{CHSH}_{ij} \nonumber
\end{align}
Then using the monogamy relations of each small piece, we arrive at the required result.\qed
\end{pf}
\end{corollary*}

\subsection{Monogamy of entropic nonlocality}
Suppose Alice is running entropic $m_i$-measurement entropic Bell experiment with $k$ Bob simultaneously, and Alice shares at least two measurements for all experiments, then we have the monogamy relations:
\begin{equation}\label{}
\mathcal{B}^{m_1}+\cdots +\mathcal{B}^{m_k}\overset{Q}{\leq}0,
\end{equation}
where $\mathcal{B}^{m_1}=H(A^i_0|B^i_{m_i})-[H(A^i_0|B^i_{0})+H(B^i_{0}|A^i_1)+\cdots +H(A^i_{m_i}|B^i_{m_i})]$, $A^i_j$ and $B^i_l$ are measurements implemented by Alice and $i$-th Bob.

Actually, this is a direct corollary of theorem \ref{thm:monogamy} and corollary \ref{cor:mono}, since the party graph $G(\mathcal{P})$ is a star graph with $k$ edges.

\subsection{Monogamy of entropic contextuality}
For BKS scenario $(\mathcal{M},\mathcal{C})$ with $\mathcal{M}=\{A_1,\cdots,A_n\}$, $\mathcal{C}=\mathcal{C}(\mathcal{M})=\{\{A_1,A_2\},\cdots,\{A_n,A_1\}\}$ and BKS scenario $(\mathcal{M}',\mathcal{C})'$ with $\mathcal{M}=\{A'_1,\cdots,A'_m\}$, $\mathcal{C}'=\mathcal{C}(\mathcal{M})=\{\{A'_1,A'_2\},\cdots,\{A'_m,A'_1\}\}$, where $A_2=A'_2$, $A_n=A'_m$, $A_1$ is compatible with all $A'_3,\cdots,A'_{m-1}$ and $A'_1$ is compatible with all $A_3,\cdots,A_{n-1}$, we must to give the needed basic entropic inequalities. For measurement set $\{A'_1,A_2,\cdots,A_n\}$ we have
\begin{align*}\label{}
   &H(A'_1A_n)\leq H(A'_1A_2\cdots A_n)  \\
   &H(A'_1|A_2\cdots A_n)\leq H(A'_1|A_2) \\
   &H(A_2|A_3\cdots A_n)\leq H(A_2|A_3)\\
   &\vdots \\
   &H(A_{n-2}|A_{n-1}A_n)\leq H(A_{n-2}|A_{n-1})
\end{align*}
For measurement set $\{A_1,A'_2,\cdots,A'_n\}$ we have
\begin{align*}\label{}
   &H(A_1A'_m)\leq H(A_1A'_2\cdots A'_m)  \\
   &H(A_1|A'_2\cdots A'_m)\leq H(A_1|A'_2) \\
   &H(A'_2|A'_3\cdots A'_m)\leq H(A'_2|A'_3)\\
   &\vdots \\
   &H(A'_{n-2}|A'_{m-1}A'_m)\leq H(A'_{m-2}|A'_{m-1})
\end{align*}
Adding all these element inequalities together and using the chain rule of conditional entropy, we obtain the inequality
\begin{multline}\label{}
H(A_1|A_n)-[H(A_1|A_2)+\cdots+H(A_{n-1}|A_n)]+ H(A'_1|A'_n)  \nonumber \\
-[H(A'_1|A'_2)+\cdots+H(A'_{n-1}|A'_n)]\overset{Q}{\leq} 0,
\end{multline}
which is nothing more than the monogamy relation of contextuality:
\begin{equation}\label{}
\mathcal{I}^{cycle}_{\mathcal{M}}+\mathcal{I}^{cycle}_{\mathcal{M}'}\overset{Q}{\leq}0.
\end{equation}

\subsection{Monogamy of entropic nonlocality and entropic contextuality}
Suppose that Alice and Bob are running a entropic nonlocality test experiment with $\mathcal{M}=\{A_0,A_2,B_0,B_1\}$ and $\mathcal{C}(\mathcal{M})=\{\{A_0,B_0\},\{A_0,B_1\},\{A_2,B_0\},\{A_2,B_1\}\}$ where $A_0,A_2$ are implemented by Alice and $B_0,B_1$ are implemented by Bob. If Alice runs an entropic contextuality text with $\mathcal{M}'=\{A_0,A_1,\cdots,A_{4}\}$ and $\mathcal{C}(\mathcal{M}')=\{\{A_0,A_1\},\cdots,\{A_{4},A_0\}\}$ simultaneously, then two experiments are monogamous.

For measurement set $\{B_1,A_2,\cdots,A_0\}$ we have
\begin{align*}\label{}
   &H(B_1A_0)\leq H(B_1A_2\cdots A_n)  \\
   &H(B_1|A_2\cdots A_0)\leq H(B_1|A_2) \\
   &H(A_2|A_3\cdots A_0)\leq H(A_2|A_3)\\
   &H(A_{3}|A_{4}A_0)\leq H(A_{3}|A_{4})
\end{align*}
For measurement set $\{A_0,A_1,A_2,B_0\}$ we have
\begin{align*}\label{}
   &H(A_1A_0)\leq H(A_1B_0A_2A_0)  \\
   &H(A_1|A_2B_0A_0)\leq H(A_1|A_2) \\
   &H(A_2|B_0A_0)\leq H(A_2|B_0)\\
\end{align*}
Adding all these element inequalities together and using the chain rule of conditional entropy, we obtain the inequality
\begin{multline}\label{}
H(A_1|A_0)-[H(A_1|A_2)+\cdots+H(A_{4}|A_0)]+ H(B_1|A_0)  \nonumber \\
-[H(B_1|A_2)+H(A_2|B_0)+H(B_{0}|A_0)]\overset{Q}{\leq} 0,
\end{multline}
which is nothing more than the monogamy relation of entropic contextuality and entropic nonlocality:
\begin{equation}\label{}
\mathcal{B}^{CHSH}_{\mathcal{M}}+\mathcal{I}^{KCBS}_{\mathcal{M}'}\overset{Q}{\leq}0.
\end{equation}

\bibliographystyle{apsrev4-1-title}

\end{document}